\documentclass[3p,times]{elsarticle}

\usepackage{ecrc}
\usepackage[final]{epsfig}

\volume{00}

\firstpage{1}

\journalname{Nuclear Physics A}

\runauth{T. Renk}

\jid{nupha}

\usepackage{amssymb}

\usepackage[figuresright]{rotating}

\begin{document}

\begin{frontmatter}



\dochead{}

\title{Modelling jet quenching}


\author{Thorsten Renk}

\address{Department of Physics, P.O. Box 35, FI-40014 University of Jyv\"askyl\"a, Finland and \\Helsinki Institute of Physics, P.O. Box 64, FI-00014 University of Helsinki, Finland}

\begin{abstract}
High $P_T$ measurements of hard hadrons or jets at RHIC and LHC appear contradictory and in some cases counter-intuitive, but upon closer investigation they represent a coherent picture of jet-medium interaction physics which can be established with systematic comparisons of models against a large body of data. This picture is consistent with a perturbative QCD mechanism and does not require exotic assumptions. This overview outlines how several key measurements each partially constrain shower-medium interaction physics and how from the sum of those the outlines of the mechanism of jet quenching can be deduced. Most current jet results from LHC can be naturally understood in this picture. A short summary of what can be established about the nature of parton-medium interaction with current data is given in the end.
\end{abstract}

\begin{keyword}
jet quenching 


\end{keyword}

\end{frontmatter}


\section{Introduction}
\label{S-Intro}

High transverse momentum ($P_T$) processes in Quantum Chromodynamics (QCD)  have been suggested as a probe to investigate the physics of ultrarelativistic heavy-ion (A-A) collisions more than a decade ago and have become a cornerstone of the experimental heavy-ion program both at RHIC and LHC. Yet, compared with for instance the modelling of the bulk medium evolution using hydrodynamics where an era of precision fits to differential data has begun (see e.g. \cite{VISHNU}), theoretical progress in the physics of hard probes is slow.

Partially, this can be attributed to the structure of the problem: Hard probes are at the same time sensitive to the  microscopical degrees of freedom of the medium and the  macroscopical medium density distribution and evolution. Thus high $P_T$ observables have the capability to potentially do both tomography and transport coefficient measurements in the medium, i.e. to constrain its microscopical and macroscopical dynamics. While the microscopical physics of the medium and hence the basic structure of the parton-medium interaction is currently in detail unknown, the density evolution (in terms of fluid-dynamical modelling) is constrained by bulk data, but not unambiguously determined. As a result, the description of any single observable in terms of a parton-medium interaction scenario and a medium evolution model is usually far from unique and does not allow a firm conclusion with regard to either question.

However, systematic multi-observable studies of a matrix of parton-medium interaction and medium evolution models can reduce the allowed model space significantly, resulting in a class of models which shows good consistency across RHIC and LHC observables.

\section{A classification of models}

\subsection{The vacuum baseline}

In vacuum, at the core of the computation of hard processes in perturbative QCD (pQCD) is the factorization of the hard, short-distance interaction part and the long-distance physics of initial state parton distributions and final state parton showers. In a medium, QCD factorization is not strictly proven, but remains a plausible working assumption. A sufficiently large separation of hard and thermal momentum scale then ensures that the hard process itself receives no medium corrections and can serve as a 'standard-candle', producing high $p_T$ partons at a known and calculable rate.

The partons emerging from the hard process are typically characterized by a large virtuality scale $Q$. If one takes the uncertainty relation to estimate the timescale at which intermediate shower states at lower virtuality can be formed, one finds $\tau \sim E/Q^2$ with $E$ the virtual parton energy. Comparing this scale with typical medium lifetimes $O(10)$ fm, one finds that for typical RHIC and LHC kinematics a significant part of the partonic shower evolution (i.e. with $Q > \Lambda_{QCD}, m_h$) takes place in the medium, with the shower being gradually boosted out of the medium at top LHC energies and on the other hand hadrons with large mass $m_h$ being produced in the medium. However, over a large kinematical range and for sufficiently light hadrons it can reasonably be assumed that the medium modification largely concerns the evolution of a parton shower with subsequent hadronization as in vacuum.

As a baseline example for in-medium shower modelling, consider a virtuality-ordered parton shower such as implemented in the PYSHOW algorithm \cite{PYSHOW}. This is treated as an interated series of splittings of a parent into two daughter partons $a \rightarrow bc$ where $E_b = z E_a$ and $E_c = (1-z) E_a$ and where the virtuality of partons in terms of $t = \ln Q^2/\Lambda_{QCD}^2$ decreases in each branching. The splitting probability is given by

\begin{displaymath}
dP_a = \sum_{b,c} \frac{\alpha_s(t)}{2\pi} P_{a\rightarrow bc}(z) dt dz
\end{displaymath}.

where $P_{a\rightarrow bc}(z)$ are perturbatively calculable objects, the so-called splitting kernels. The splittings cease when the parton virtuality reaches a non-perturbative lower scale $O(1)$ GeV at which point a hadronization model needs to be used. In this way, the original high virtuality of the shower initiator is converted into the transverse momentum distribution of the final shower partons (and the initial virtuality sets a kinematic bound for observables like jet shapes).

An important thing to note is that the  splitting kernels $P_{a\rightarrow bc}(z)$ are scale invariant, i.e. they do not depend on an absolute momentum scale. As a result, the fragmentation functions generated by a series of splittings is self-similar (the momentum distribution of hadrons in a subjet looks almost the same as the distribution in the whole jet when plotted as a function of $z$) and up to logarithmic corrections has no strong energy dependence.

Estimating the timescale associated with the initial, hard branchings which largely determine jet shape and subjet structure, one finds $\tau \sim 0.01$ fm, i.e. well before any medium can be formed. This gives rise to the expectation that subject structure and jet shape at high $P_T$ should not be very sensitive to medium modifications.

\subsection{In-medium showers}

There are two main processes by which this pattern can be modified in a medium: 1) direct loss of energy (by elastic collisions, drag, \dots) into the medium as parametrized by a transport coefficient $\hat{e}$ and 2) enhanced soft gluon radiation by medium-induced virtuality as parametrized by a transport coefficient $\hat{q}$. The first mechanism is in general incoherent and the energy is lost into the medium, the second mechanism involves a de-coherence time for the radiated gluon and transfers energy from the leading parton into an increased number of subleading shower partons. Note that this distinction can be made on average on the basis of transport coefficients, but not for a single interaction graph with the medium which may involve both direct and radiative energy loss. A recently suggested additional mechanism is the modification of color flow by the medium, leading to a modification of the hadronic shower evolution \cite{ColorFlow}.

Models may now be classified based on what part of the dynamics they include, how they treat the medium and how they solve the shower evolution equations. The most complete approach are models which simulate the whole in-medium shower evolution such as JEWEL \cite{JEWEL},  YaJEM \cite{YaJEM1,YaJEM2}, Q-PYTHIA \cite{Q-PYTHIA} or resummed higher twist (HT) \cite{resHT}. However, if one is only interested in observables which probe the leading shower fragments, such as single inclusive high $P_T$ hadron suppression, which probe dominantly showers in which a large fraction of the energy flows through a single parton, then leading parton energy loss modelling in which the medium evolution is cast into an energy loss probability $P(\Delta E)$ before vacuum fragmentation is applied is sufficient. Models of this type include the ASW \cite{ASW-1}, AMY \cite{AMY-1,AMY-2} or WHDG \cite{WHDG} frameworks. Finally, in hybrid approaches a vacuum shower is computed down to some scale, then the produced partons are put on-shell and each propagated through the medium using a leading-parton energy loss model. An example for hybrid modelling is MARTINI \cite{MARTINI}.

Of considerable interest is how the medium is included into the modelling process. A number of models come with an explicit treatment of medium partons and their interaction with shower partons, typically as a thermal gas of quark and gluon quasiparticles \cite{JEWEL,AMY-1,AMY-2,MARTINI}. Such a description requires that the medium is, at least to some degree, perturbatively tractable, which is not in line with the basic assumption underlying almost ideal hydrodynamical modelling which assumes that the medium is strongly coupled and the mean free path essentially vanishes. Usually an explicit treatment of interactions with medium partons breaks the scale invariance of the fragmentation function around the momentum scale of the medium partons and below, while it largely remains intact above. In addition, the medium provides additional kinematic phase space for transverse shower broadening.

A different approximation is to cast all medium effects into a modification of the splitting probability, in essence replacing $P_{a\rightarrow bc}(z) \rightarrow P'_{a\rightarrow bc}(z)$ with the detailed form of $P'$ motivated by a specific physics scenario as done e.g. in Q-PYTHIA \cite{Q-PYTHIA}. Such a prescription explicitly conserves energy and momentum inside the shower, i.e. there is no additional phase space for transverse broadening. It also perserves  scale invariance of the fragmentation function, but it alters its shape to a different functional form.

Finally, the medium can be (without any explicit reference to its degrees of freedom) appear via transport coefficients such as $\hat{q}, \hat{e}$ and alter the kinematics of propagating partons as implemented in YaJEM \cite{YaJEM1,YaJEM2}. This again breaks scale invariance of the fragmentation function around the medium momentum scale and below and leads to additional phase space for transverse broadening.

There are two main strategies to solve the shower evolution equations --- analytical and Mote-Carlo (MC) techniques. In general, analytical techniques get exact treatment of quantum interference effects while they often rely on kinematic approximations such as eikonal propagation or infinite parent parton energy which violate energy-momentum conservation. In contrast, MC frameworks usually have exact kinematics, but have to resort to phenomenological prescriptions for interference effects.

\begin{table}[htb]
\begin{center}
\begin{tabular}{|l|lll|}
\hline
& medium modifies kinematics & probabilistic picture & explicit medium partons\\
\hline
MC shower & YaJEM \cite{YaJEM1,YaJEM2} & Q-PYTHIA \cite{Q-PYTHIA}, BW \cite{BW} & JEWEL \cite{JEWEL}\\
analytical shower & --- & HT resummed \cite{resHT} & ---\\
MC hybrid &--- & PYQUEN \cite{PYQUEN} & MARTINI \cite{MARTINI}\\
analytical energy loss & --- & ASW \cite{ASW-1}, HT \cite{HT} & AMY \cite{AMY-1,AMY-2}, WHDG \cite{WHDG}\\ 
\hline
\end{tabular}
\caption{\label{T-Models}A summary table of a rough classification of several well-known jet quenching models.} 
\end{center}
\end{table}

These different approaches to solving the in-medium shower evolution equations are summarized for several well-known models in table \ref{T-Models}. Model results are also quantitatively sensitive to implementation details such as cutoffs or the choice of light-cone vs. energy splitting. However, these implementation specifics do not seem to lead to qualitative changes of the model results. A very instructive review of such effects can be found in \cite{Brick}.

\section{Physics assumptions in jet quenching models}

In addition to implementing different approximations and solution strategies, models also show genuine differences with regard to assumptions about the physics underlying parton-medium interaction, and this leads to various testable consequences. In particular, the pathlength dependence of the mean energy loss $\Delta E$ of a parton traversing a constant medium is parametrically different dependent on what one assumes to be the relevant physics (note that $\Delta E$ is not as such a very well-defined concept for an in-medium shower, but it can be extracted by tagging a quark flavour as done e.g. in \cite{YaJEM2}).

For any incoherent mechanism, the mean energy loss largely tracks the number of scatterings, which in a constant medium are proportional to the length, i.e. $\Delta E \sim L$. For radiative processes, there is a decoherence time involved and hence generic arguments can be made (e.g. \cite{QuenchingWeights}) that $\Delta E \sim L^2$. It is however known that this holds only for sufficiently large parent parton energy and finite energy corrections quickly change this to a linear dependence \cite{Korinna-LPM,YaJEM-D,Caron-Huot}. In strong coupling scenarios motivated by AdS/CFT approaches to the medium dynamics, a longitudinal drag rather than transverse random momentum transfer lead to induced radiation, which changes the above argument dimensionally, leading to $\Delta E \sim L^3$ \cite{AdS}, to which finite energy correction have so far not been obtained, but would be expected to weaken the dependence.

This discussion is somewhat correlated to the way the medium is implemented in models as discussed in the previous section, as models which include explicit interactions with perturbative medium quasiparticles typically find ~50\% elastic (incoherent) energy loss for reasonable values of the strong coupling $\alpha_s \approx 0.3$ \cite{WHDG,EMC}. In contrast, many older leading parton radiative energy loss (e.g. \cite{ASW-1,HT}) start from the assumption of static scattering centers in the medium and hence do not have any incoherent component.

Yet a different source of pathlength dependence is the idea that the minimum virtuality scale down to which a shower can evolve in-medium should be constrained by uncertainty arguments to $Q_0 = \sqrt{E/L}$ \cite{resHT}. This leads to a non-linear pathlength as well as additional energy dependence of the mean energy loss \cite{YaJEM-D}.

It is important to note that in a real fluid-dynamical background the actual pathlength dependence is drastically changed by the spacetime evolution of the medium density. In this sense, expressions like 'quadratic pathlength dependence' are to be understood as labels what a model would do if applied to a constant medium, not as descriptions of what is experimentally measurable. In the following, we aim to establish experimental tests sensitive to these different physics assumptions and implementation differences such that the data can be used to discriminate between models.

\section{Single hadron suppression}

\begin{figure}[htb]
\begin{center}
\raisebox{5cm}{\epsfig{file=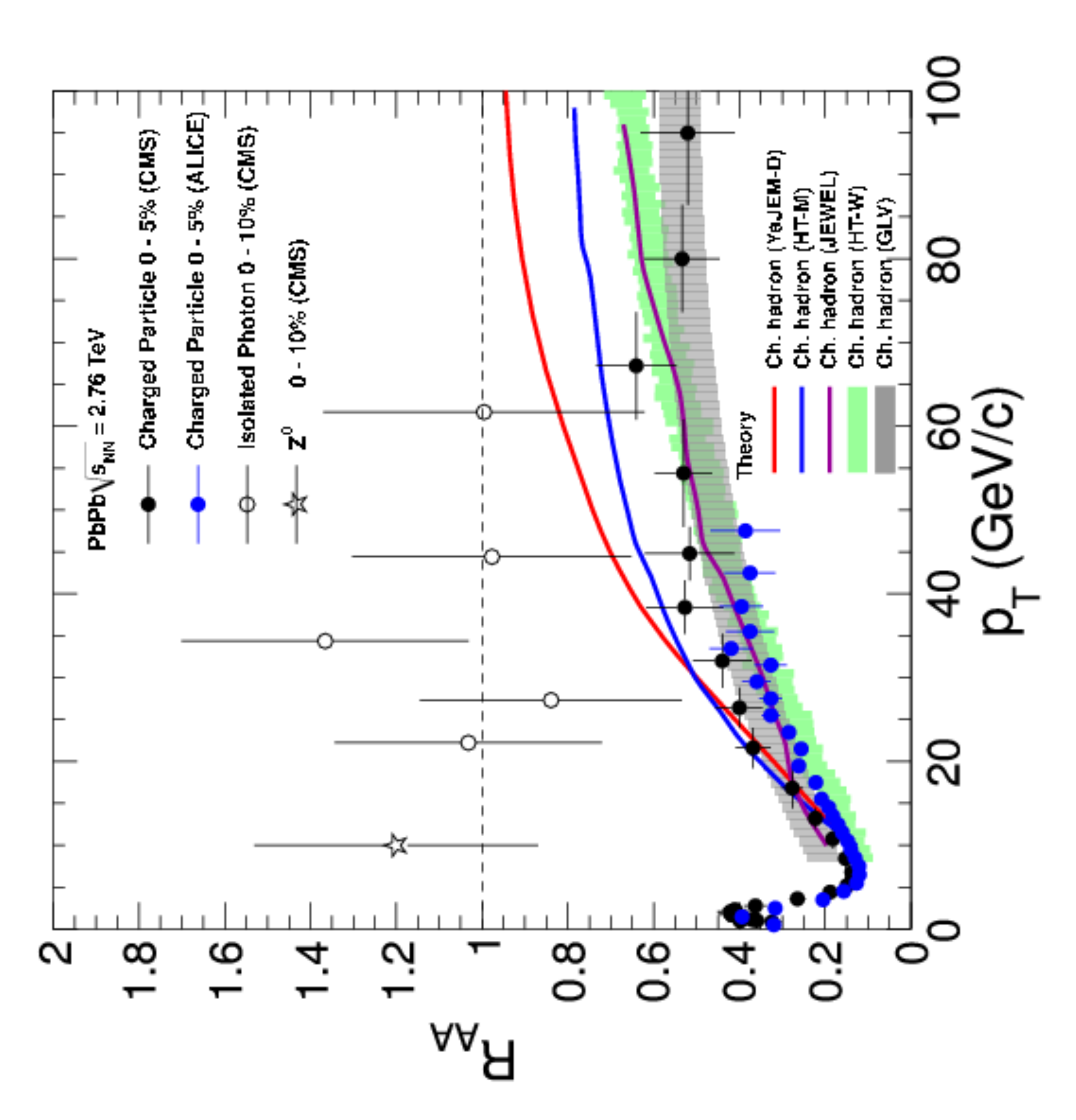, width=5cm, angle=-90}} \epsfig{file=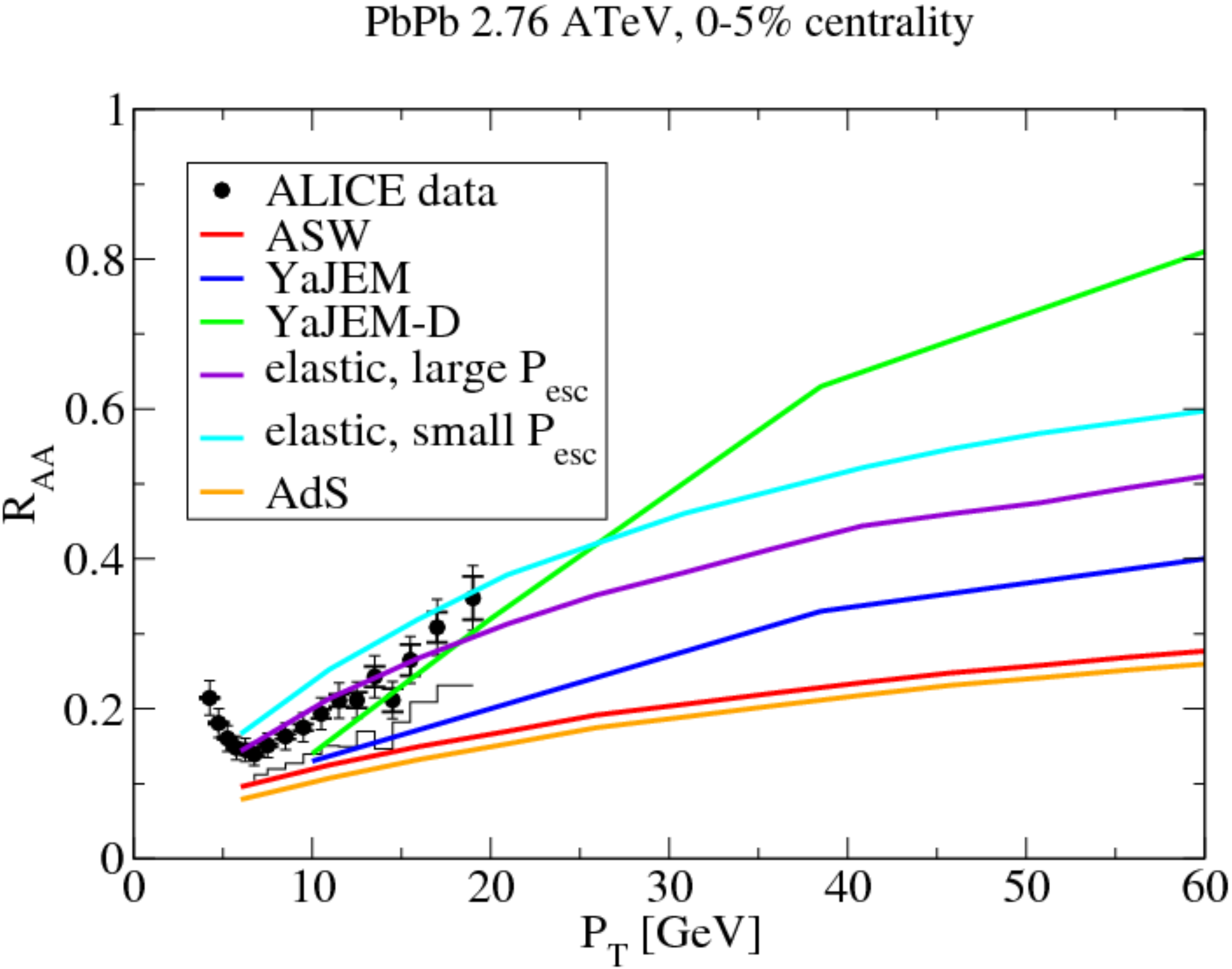, width=6.7cm}
\end{center}
\caption{\label{F-RAA-LHC}Pre- and postdictions for the nuclear suppression factor $R_{AA}$ in 2.76 ATeV central Pb-Pb collisions comparison with data from CMS \cite{CMS-RAA} and ALICE \cite{ALICE-RAA} in various models \cite{RAA_LHC, RAA_LHC_WHDG, RAA_LHC_HT, RAA_LHC_JEWEL}.}
\end{figure}

In general, the $P_T$ dependence of the single inclusive hadron suppression factor $R_{AA}$ is more driven by generic pQCD effects such as the functional form of the produced parton momentum spectrum than by jet quenching model specific effects (see \cite{ElossPhysics} for a discussion). As a result, within the uncertainty associated with the choice of the bulk medium evolution model, models tuned to RHIC data tend to extrapolate reasonably well to LHC conditions (see Fig.~\ref{F-RAA-LHC}).

The important exception is the strong coupling scenario denoted as AdS which overquenches significantly. This can be understood as follows: Due to dimensional reasons, the $L^3$ pathlength dependence requires a scaling of the quenching effect with the medium temperature $\sim T^4$ whereas all other scenarios scale with the approximate medium density $\sim T^3$. Such a strong increase of the quenching power is however not indicated by the data, essentially ruling out this particular application of AdS/CFT ideas to jet quenching.

$R_{AA}$ is, however, not expected to be a very sensitive probe \cite{ElossPhysics} --- due to the need to average over the full medium geometry and to fold with the pQCD parton spectra, only a narrow region of the medium-modified fragmentation function (MMFF) is probed by the observable, and thus models predicting a very different MMFF can result in nearly the same $R_{AA}$. One way to constain models better is to study the dependence of $R_{AA}$ on the angle of the hadron with the reaction plane $\phi$. In this way, the pathlength dependence of the parton-medium interaction model \emph{in combination with a particular medium evolution model} is tested. Such investigations have established two crucial insights \cite{ElossPhysics,JetHydSys}:

\begin{itemize}
\item There is about a factor two uncertainty associated with the choice of the medium model provided that the medium model is a fluid-dynamical model constrained by bulk observables. One implication is that $R_{AA}(\phi)$ has the capability to discriminate between fluid-dynamical models, i.e. it is a true tomographical measurement. However, this result also indicates that any results obtained with a schematical medium evolution model not constrained by bulk data (such as the Bjorken cylinder) should be disregarded, as they have easily a factor 10 systematic uncertainty.

\item Within the factor two uncertainty, most scenarios describe the data reasonably well. The one exception which fails by a huge margin (factor 6) is a linear pathlength dependence \cite{EMCRP}. Based on comparison with the data, any component with linear pathlength dependence must be smaller than about 10\% \cite{ElossPhysics,EMCRP,ElPhenom}. This does not only disfavour scenarios with a 50\% elastic energy loss component, but also any radiative scenario with finite energy corrections. The only realistic pathlength-dependence generating option viable with the combined data is hence the prescription for the minimum in-medium virtuality $Q_0 = \sqrt{E/L}$ as suggested in \cite{resHT} (both radiative energy loss from an infinite energy parton and an AdS $L^3$ dependence would work with the data, but the first is not a sufficiently realistic scenario and the second alternative does not agree well with the extrapolation from RHIC to LHC as discussed above).
\end{itemize}

\section{Hard dihadron correlations}

The normalized dihadron away side correlation strength $I_{AA}$ is an interesting observable, since the requirement of a hard near side trigger adds a series of biases on parton type and kinematics while the away side fragmentation pattern has no additional bias \cite{Dihadrons}. As a result, the MMFF is probed in more detail in such measurements.

This allows to study the redistribution of longitudinal momentum inside a shower in quite some detail. The two main physics scenarios outlined above expect rather different behaviour: In a direct, elastic energy transfer into the medium, longitudinal momentum lost from the jet is carried by the medium at thermal momentum scales (i.e. not in hard modes). In contrast, in a radiative energy loss scenario, longitudinal momentum is carried by additional soft gluon radiation, part of which is still 'hard' as compared to a thermal momentum scale. Dihadron correlation are able to discriminate between these patterns.

\begin{figure}[htb]
\begin{center}
\epsfig{file=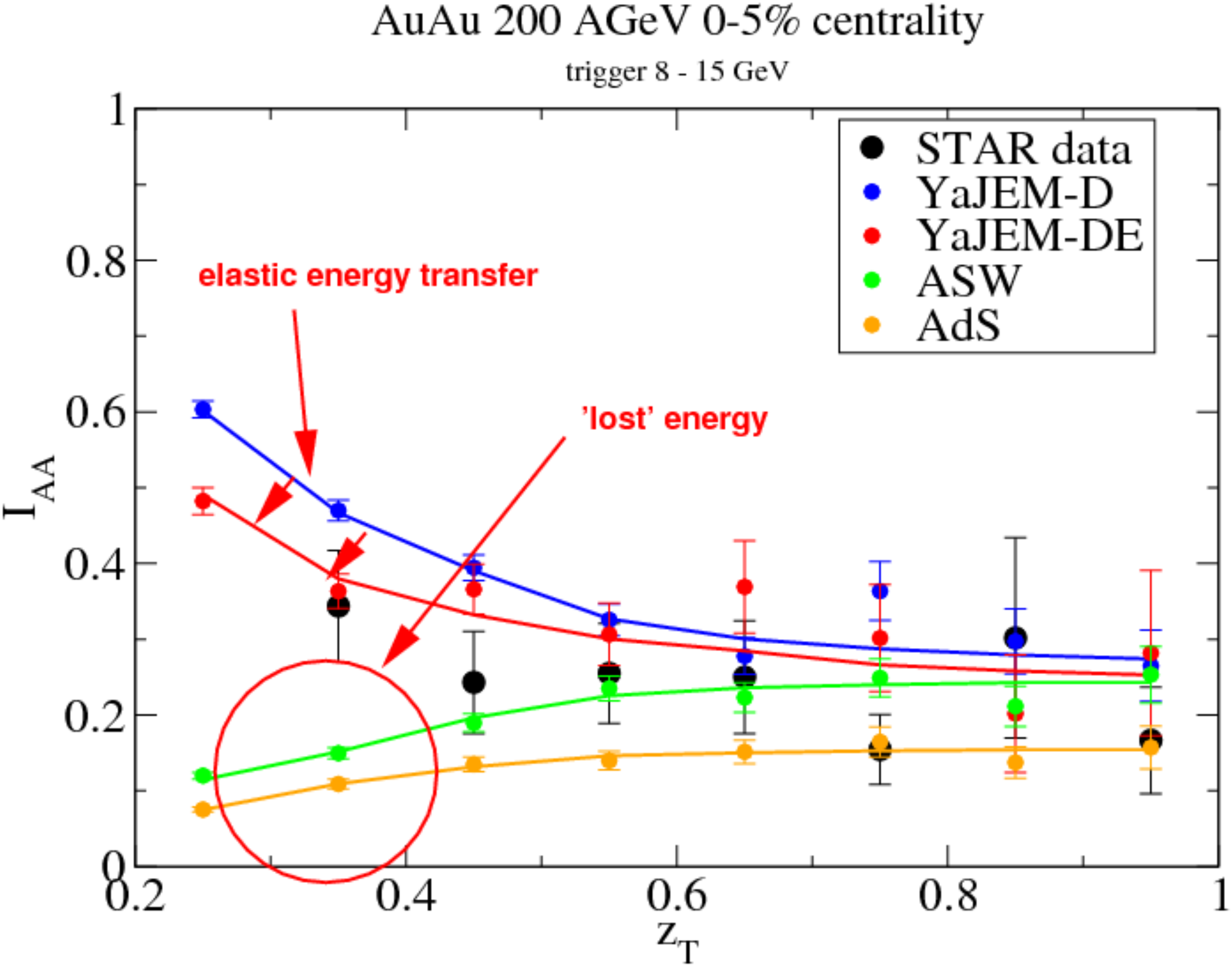, width=6cm}
\end{center}
\caption{\label{F-IAA}Away side normalized dihadron correlation strength $I_{AA}$ as compared with leading parton energy loss models and in-medium shower models \cite{Dihadrons,DihadronsEl}.}
\end{figure}

This is illustrated in Fig.~\ref{F-IAA} where the away side correlation strength as a function of $z_T = E_{hadron}/E_{trigger}$ for central 200 AGeV Au-Au collisions and a 8-15 GeV trigger is shown. In leading parton energy loss calculations, energy is implicitly assumed to be transferred into the medium and to reappear at thermal momentum scales (below the $z_T$ range of the measurement). As a result, the curves bend downward at low $z_T$ which is not in agreement with the trend seen in the data. In contrast, in in-medium shower pictures where the enhanced production of subleading jet fragments is explicitly treated, the curves reflect the enhanced soft gluon production by bending upward.

However, under the assumption that all energy re-appears in soft gluon production, the computation overshoots the data. The tend seen in the data can only be reproduced under the assumption that about 10\% of the energy is directly tranferred into the soft medium \cite{DihadronsEl}. This lower limit for the direct elastic energy transfer agrees nicely with the upper limit of the same order as obtained by a study of pathlength dependent observables.
Thus, this observable clearly shows the limits of leading parton energy loss modelling and also disfavours all scenarios in which there is no possibility of direct energy transfer into the medium.

\section{Clustering and suppression of jets}

The clustering of hadron showers into jets by means of a clustering algorithm is designed to suppress the dependence of observables to soft physics such as hadronization and to provide a good comparison point between hard pQCD calculations and measurements without the need to model physics close to $\Lambda_{QCD}$. As a result, clustering suppresses many effects of a medium modifiaction since these take place at a scale $T \sim \Lambda_{QCD}$. To provide a concrete example, while the medium-induced emission of an almost collinear gluon leads to a modification of the leading hadron spectrum since energy has been taken from the leading parton, it does not lead to a modification of the jet spectrum since the emitted gluon is clustered back into the jet. Thus, unless distributions of hadrons inside the clustered jets are considered, jets are significantly less sensitive to medium modifications than single hadrons \cite{DijetsATLAS}.

There is, however, a two stage mechanism which is able to suppress jets: While hard partons with $p_T \gg T$ are kinematically very robust against interactions with the medium and cannot easily be scattered out of the jet, the interaction can nevertheless induce the emission of additional soft gluons at a thermal energy scale. Such soft gluons however are not kinematically robust and can be scattered easily out of the jet cone, leading to jet energy loss. Since gluons at a thermal scale are indistinguishable from medium gluons, the energy radiated into these modes is quickly thermalized and flows in hydrodynamical excitations to large angles. This is thus a very generic mechanism independent of the specific in-medium shower physics. The characteristics of the parton-medium interaction are then only apparent from semi-soft gluons above the thermal scale. Such a scenario is well in line with jet quenching as characterized by RHIC observables discussed before.

This mechanism of jet suppression has recently been denoted as \emph{frequency collimation} \cite{Fcollim} but has been implemented into MC and observed earlier in \cite{JetShapes}. Since it relies on extra available phase space for rescattered soft gluons, it can not be observed in purely probabilistic approximations of the jet-medium interaction.

\section{Jet observables}

\begin{figure}[htb]
\begin{center}
\epsfig{file=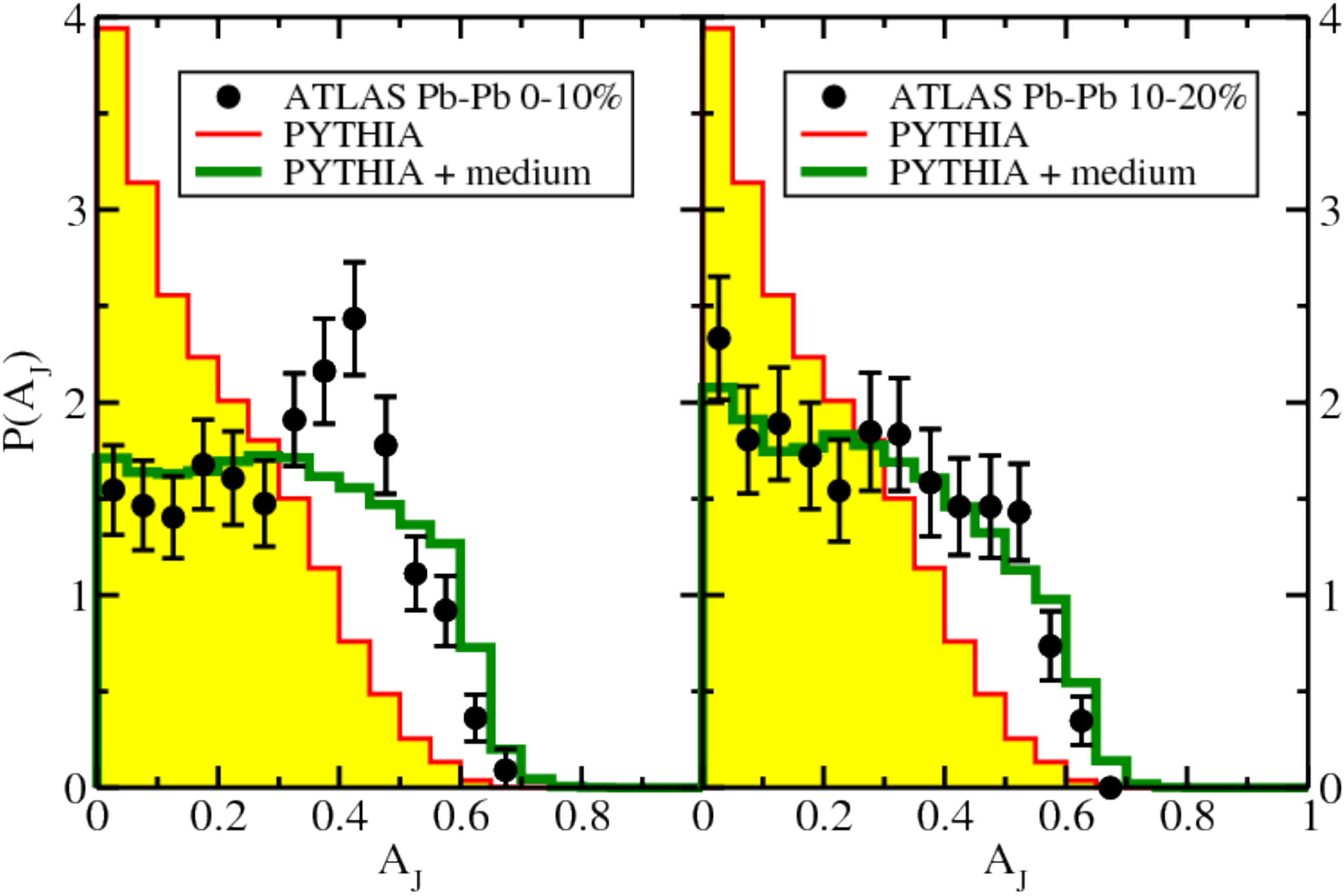, width=7cm}\epsfig{file=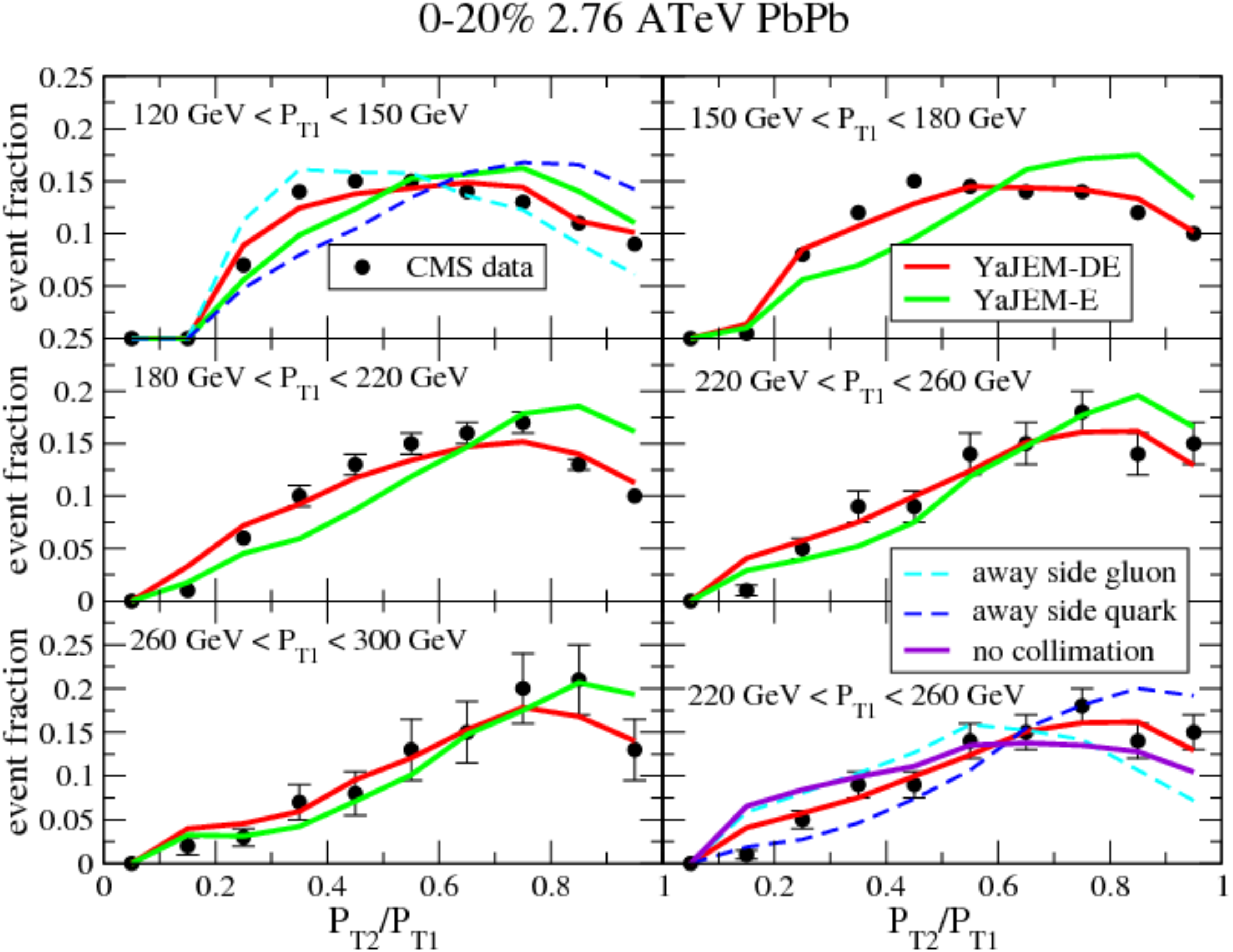, width=6cm}
\end{center}
\caption{\label{F-A_J}Dijet asymmetry as measured by ATLAS \cite{ATLAS} and CMS \cite{CMS}  as compared with medium induced radiative energy loss scenarios \cite{A_J_Qin,Dijet_Edep}}
\end{figure}

The first LHC observable involving reconstructed jets has been the dijet energy imbalance  
$A_J = \frac{E_{T1} - E_{T2}}{E_{T1}+E_{T2}}$ or simply the ratio $E_{T2}/E_{T1}$ of reconstructed jet energies $E_{T1}$ on the near side and $E_{T2}$ on the away side \cite{ATLAS,CMS}. For the reasons mentioned above, clustering largely removes the sensitivity to specific characteristics of the jet quenching model, and as a result the data can be described in various models making somewhat different assumptions about the dynamics of parton-medium interaction and medium geometry \cite{DijetsATLAS,A_J_Qin,A_J_Vitev,A_J_Young}. The trigger energy dependence of the imbalance is somewhat more constraining, it probes the medium-induced broadening of the jet \cite{Dijet_Edep} (see Fig.~\ref{F-A_J}).

A greater challenge to models is posed by the CMS observation that the longitudinal momentum distribution of reconstructed jets, when plotted as a fraction of the reconstructed jet energy, appears unchanged from vacuum even for highly imbalanced events \cite{FPattern}. This is a highly unexpected finding for any model which implements the interaction with the medium probabilistically and hence uses a modified set of splitting kernels $P_{a\rightarrow b,c}'(z)$ to build the fragmentation function, as in such a model the MMFF would be self-similar, but not show the same functional form as in vacuum.
It is however a natural outcome for models which break scale invariance only below a fixed momentum scale $O(T)$, since in such frameworks the splitting kernels generate a self-similar fragmentation function with the same functional form as in vacuum above the breaking scale. The flattening of $I_{AA}(z_T)$ for $z_T > 0.5$ in Fig.~\ref{F-IAA} is a manifestation of the same dynamics. From the RHIC data, significant deviations from the vacuum dynamics are expected to occur around 2-3 GeV, below the range of the CMS measurement. Thus, the CMS finding of an apparently unmodified fragmentation pattern strongly disfavours an probabilistic implementation of medium effects in the splitting kernel, but is not unexpected for other models, indeed almost unmodified jet properties above a momentum cut of 4 GeV with significant modifications below had been predicted in \cite{JetShapes}.

\section{Summary}

Our knowledge of the viability of various models, given the available data and aking use of the full uncertainty range given by the choice of the hydrodynamical evolution model can be summarized as in Table~\ref{T-summary}: 

\begin{table}[htb]
\begin{tabular}{|l|cccccc|}
\hline
& $R_{AA}(\phi)$@RHIC & $R_{AA}$@LHC ($P_T$)& $I_{AA}$@RHIC & $I_{AA}$@LHC &  $A_J$ & $E_{T2}/E_{T1} (E_{trig})$ \\
\hline 
elastic &  {fails} &  {works} &  {fails} & fails & works & fails\\
elMC &  {fails} &  {fails} &  {fails} & not tested & not tested & not tested \\
ASW &    {works} &  {fails} &  {marginal} & works & N/A & N/A \\
AdS &    {works} &  {fails} &  {marginal}  & works & N/A & N/A\\
YaJEM &  {fails} &  {fails} &  {fails} & fails & works & works\\
YaJEM-D &  {works} &  {works} &  {marginal} & marginal & works & works \\
YaJEM-DE &  {works} &  {works} &  {works}  & works & works & works\\
\hline
\end{tabular}
\caption{\label{T-summary}Viability of different parton-medium interaction models tuned to the $P_T$ dependence of $R_{AA}$ in 200 AGeV Au-Au collisions given various data sets under the assumption that the best possible hydrodynamical evolution scenario is chosen. The various labels refer to: elastic \cite{ElPhenom}, elMC \cite{EMCRP}, ASW \cite{QuenchingWeights}, AdS \cite{AdS}, YaJEM \cite{YaJEM1,YaJEM2}, YaJEM-D \cite{YaJEM-D}, YaJEM-DE \cite{DihadronsEl}.}
\end{table}

It is immediately evident that the combination of all observables cuts the available model space down much more than any single observable. Most of the constraints are provided by the combination of the $P_T$ dependence of $R_{AA}$ at LHC which probes the dynamics of leading partons in combination with a dihadron correlation $I_{AA}$ which is a probe of subleading fragment dynamics. Current jet measurements do not have strong constraining power beyond this, however they do highlight the need to go beyond a purely probabilistic implementation of medium modification and also provide constraints for transverse dynamics of showers. A number of conclusions with regard to the nature of jet-medium interaction can be drawn from this:

\begin{itemize}
\item The combined data is compatible with a fairly standard pQCD picture of medium-induced radiation and a subleading component of elastic interactions leading to energy transfer into the medium
\item There is no evidence for exotic scenarios, the data neither indicate an $L^3$ dependence of energy loss as suggested by some variants of AdS/CFT inspired models nor a modification of the hadronization stage. 
\item Data from RHIC and LHC consistently indicate that some moderate amount of energy is transferred from the hard parton directly into the non-perturbative medium rather than into subleading jet fragments. However, the amount of direct energy transfer is constrained by pathlength-dependent observables and found to be significantly smaller than if the medium can be described as a near-ideal quark-gluon gas. This may imply that the observed degrees of freedom in the medium are massive or correlated quasiparticles.
\end{itemize}

To make these conclusions more quantitative, further high-statistics multi-differential measurements are needed from the experimental side, followed by systematic multi-observable studies from theory. In particular, the systematics of h-h, jet-h and $\gamma$-h correlations (with decreasing geometry bias) as a function of the reaction plane angle for various system centralities is expected to be a sensitive probe of both the parton-medium interaction mechanism and the medium density evolution. Once these questions are reliably addressed, the precise nature of the mechanism by which the energy deposited from hard partons is carried by the medium presents itself as the next goal. 





\bibliographystyle{elsarticle-num}



\end{document}